\documentclass[12pt]{article}
\usepackage{amsmath,amssymb,epsfig,bbm,capt-of,ifthen,calc}

\catcode`\à=\active \defà{\`a} \catcode`\À=\active \defÀ{\`A}
\catcode`\á=\active \defá{\'a} \catcode`\Á=\active \defÁ{\'A}
\catcode`\ä=\active \defä{\"a} \catcode`\Ä=\active \defÄ{\"A}
\catcode`\â=\active \defâ{\^a} \catcode`\Â=\active \defÂ{\^A}
\catcode`\å=\active \defå{{\aa}} \catcode`\Å=\active \defÅ{{\AA}}
\catcode`\ç=\active \defç{\c{c}} \catcode`\Ç=\active \defÇ{\c{C}}
\catcode`\è=\active \defè{\`e} \catcode`\È=\active \defÈ{\`E}
\catcode`\é=\active \defé{\'e} \catcode`\É=\active \defÉ{\'E}
\catcode`\ë=\active \defë{\"e} \catcode`\Ë=\active \defË{\"E}
\catcode`\ê=\active \defê{\^e} \catcode`\Ê=\active \defÊ{\^E}
\catcode`\ì=\active \defì{\`{\i}} \catcode`\Ì=\active \defÌ{\`{\I}}
\catcode`\í=\active \defí{\'{\i}} \catcode`\Í=\active \defÍ{\'{\I}}
\catcode`\ï=\active \defï{\"{\i}} \catcode`\Ï=\active \defÏ{\"{\I}}
\catcode`\î=\active \defî{\^{\i}} \catcode`\Î=\active \defÎ{\^{\I}}
\catcode`\ò=\active \defò{\`o} \catcode`\Ò=\active \defÒ{\`O}
\catcode`\ó=\active \defó{\'o} \catcode`\Ó=\active \defÓ{\'O}
\catcode`\ö=\active \defö{\"o} \catcode`\Ö=\active \defÖ{\"O}
\catcode`\ô=\active \defô{\^o} \catcode`\Ô=\active \defÔ{\^O}
\catcode`\ù=\active \defù{\`u} \catcode`\Ù=\active \defÙ{\`U}
\catcode`\ú=\active \defú{\'u} \catcode`\Ú=\active \defÚ{\'U}
\catcode`\ü=\active \defü{\"u} \catcode`\Ü=\active \defÜ{\"U}
\catcode`\û=\active \defû{\^u} \catcode`\Û=\active \defÛ{\^U}
\catcode`\ý=\active \defý{\'y} \catcode`\Ý=\active \defÝ{\'Y}
\catcode`\ÿ=\active \defÿ{\"y} \catcode`\˜=\active \def˜{\"Y}
\catcode`\½=\active \def½{!`}
\catcode`\¾=\active \def¾{?`}
\catcode`\ß=\active \defß{{\ss}}
\newcommand{\keywords}[1]{{{\bf Keywords: }#1}}
\newcommand{\email}[1]{{{\em Email}: {\tt #1}}}
\newcommand{\tmfloatcontents}{}
\newlength{\tmfloatwidth}
\newcommand{\tmfloat}[5]{
  \renewcommand{\tmfloatcontents}{#4}
  \setlength{\tmfloatwidth}{\widthof{\tmfloatcontents}+1in}
  \ifthenelse{\equal{#2}{small}}
    {\ifthenelse{\lengthtest{\tmfloatwidth > \linewidth}}
      {\setlength{\tmfloatwidth}{\linewidth}}{}}
    {\setlength{\tmfloatwidth}{\linewidth}}  \begin{minipage}[#1]{\tmfloatwidth}
    \begin{center}
      \tmfloatcontents
      \captionof{#3}{#5}
    \end{center}
  \end{minipage}}
\newcommand{\tmmathbf}[1]{\ensuremath{\boldsymbol{#1}}}
\newcommand{\tmscript}[1]{\text{\scriptsize $#1$}}
\newtheorem{theorem}{Theorem}
\newtheorem{varremark}{Remark}
\newenvironment{remark}{\begin{varremark}\em}{\em\end{varremark}}
\newcommand{\tmop}[1]{\ensuremath{\operatorname{#1}}}
\newcommand{\op}[1]{#1}
\newcommand{\tmem}[1]{{\em #1\/}}
\newtheorem{lemma}{Lemma}
\newenvironment{proof}{
  \noindent\textbf{Proof}\ }{\hspace*{\fill}
  \begin{math}\Box\end{math}\medskip}
\newcommand{\equallim}{\mathop{=}\limits}
\newcommand{\assign}{:=}
\newcommand{\eqnumber}{\hfill(\theequation\addtocounter{equation}{-1}\refstepcounter{equation}\addtocounter{equation}{1})}
\begin{document}

\title{A lattice model for the line tension of a sessile  drop\thanks{Preprint
CPT--2005/P.076; }} \author{Daniel
GANDOLFO\thanks{\email{gandolfo@cpt.univ-mrs.fr}}\\
\small Centre de Physique Théorique, CNRS \\[-4pt]
\small Luminy Case 907, F-13288 Marseille Cedex
\small 9, France, 
\\[-4pt]
\small and Département de Mathématiques, Université du Sud Toulon--Var, 
\\[-4pt]
\small BP 132, F--83957 La Garde, France. \and 
Lahoussine
LAANAIT\thanks{\email{laanait@yahoo.fr}}
\\
\small Ecole Normale Supérieure de Rabat, 
\small  BP 5118, Rabat, Morocco. 
\and 
Salvador
MIRACLE--SOLE\thanks{\email{miracle@cpt.univ-mrs.fr}}\\
\small Centre de Physique Théorique, CNRS \\[-4pt]
\small Luminy Case 907, F--13288 Marseille Cedex
\small 9, France. 
\and Jean RUIZ\thanks{\email{ruiz@cpt.univ-mrs.fr}}\\
\small Centre de Physique Théorique, CNRS \\[-4pt]
\small Luminy Case 907, F--13288 Marseille Cedex
\small 9, France.} 
\date{}
\maketitle

\begin{abstract}
  Within a semi--infinite thre--dimensional lattice gas model describing the
  coexistence of two phases on a substrate, we study, by cluster expansion
  techniques, the free energy (line tension) associated with the contact line
  between the two phases and the substrate. We show that this line tension, is
  given at low temperature by a convergent series whose leading term is
  negative,  and equals $0$ at zero temperature.

\end{abstract}


\section{Introduction}

Suppose that we have a drop of some matter, here called the $( + )$ or the
dense phase, over a flat substrate, also called the wall, $W$, while both are
in a medium, here called the $( - )$ or the dilute phase. Equilibrium is
obtained when the free energy of the surfaces of contact is a minimum. We have
then three different surfaces of contact, and the total free energy of the
system consists of three parts, associated to these three surfaces. A drop of
the dense phase will exist provided its own two surface tensions exceed the
surface tension between the substrate $W$ and the medium, i.e., provided that
\begin{equation}
  \tau^{\text{\textrm{w}} +} + \tau^{+ -} > \tau^{\text{\textrm{w}} -} .
  \label{pw}
\end{equation}

If equality is attained then a film of dense phase is formed, a situation
which is known as perfect, or complete wetting.

When the substance involved is anisotropic, such as a crystal, the
contribution to the total free energy of each element of area of the interface
between the dense and the dilute phases depends on its orientation. The
minimum surface free energy for a given volume of matter determines then, the
ideal form of the sessile drop at equilibrium (Fig. 1). This form is given by
the Winterbottom construction {\cite{W}}.

The above description is valid only if there is no free energy per unit
length, or line tension, associated to the line of contact of the surface of
the drop with the wall, or if the size of the drop tends to infinity. If it is
not the case, this analysis has to be revisited, see e.g. refs. {\cite{LP-W}},
{\cite{LP-GRV}}, {\cite{LP-KM}}, {\cite{LP-ABS}}, {\cite{LP-S}}.

To examine some theoretical aspects of this question in the frame of
statistical mechanical models will be the object of the present study. More
precisely, for a lattice gas model describing the coexistence of a dense and
diluted phase on a wall, we analyse the free energy of the contact line
between these phases and the wall. 

\tmfloat{h}{big}{figure}{\epsfig{file=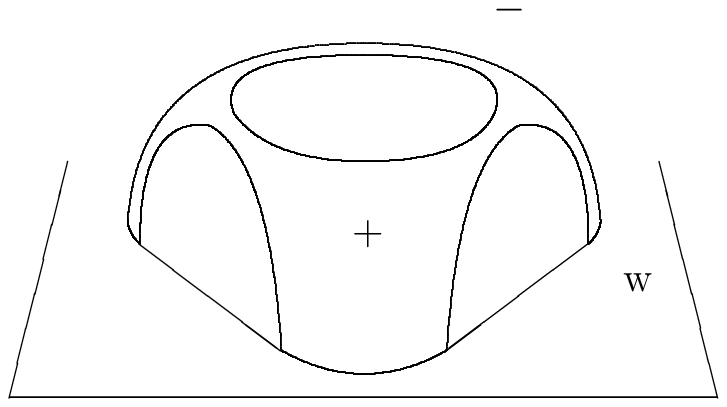}}{\small Sessile drop on a substrate}

\vspace{1cm}

For the study of this coexistence, a lattice system has been already
introduced by Fröhlich and Pfister in refs. {\cite{FPa}}, {\cite{FPb}},
{\cite{FPc}}, namely, the Ising model defined on the semi-infinite lattice
\begin{equation}
  {\mathbb{L}} = \{ i = ( i_1, i_2, i_3 ) \in \mathbb{Z}^3 : i_3 \ge
  1 \},
\end{equation}
A variable $\sigma_i$, that may take the two values $1$ and $- 1$, is
associated to each site $i \in {\mathbbm{L}}$. For notational reasons
we consider the wall as the sublattice
\begin{equation}
  W = \{ i = ( i_1, i_2, i_3 ) \in \mathbbm{Z}^3 : i_3 = 0 \},
\end{equation}
and assume that $\sigma_i = 1$ if $i \in W$. A magnetic field, $K$, is added
on the boundary sites, $i_3 = 1$, which describes the interaction with the
substrate. The strength of the nearest neighbor ferromagnetic interaction is
denoted by $J$, the bulk magnetic field is zero, and $\beta = \frac{1}{k T}$
represents the inverse temperature. The positively magnetized phase is
interpreted as the dense phase, the negatively magnetized phase as the medium.
One defines, using the grand canonical ensemble, the surface free energies
$\tau^{\text{\textrm{w}} +}$ and $\tau^{\text{\textrm{w}} -}$ in agreement
with this interpretation, see below. Analogously, $\tau^{+ -}$ is the surface
tension of the usual Ising model for an interface orthogonal to the $i_3$
axis.

Let us mention the following results of the Fröhlich and Pfister study: For
$|K| < J$, the surface tensions $\tau^{\text{\textrm{w}} +} ( \beta )$ and
$\tau^{\text{\textrm{w}} -} ( \beta )$ are analytic functions at low
temperatures, i.e., provided that $\beta ( J - |K| ) > c_0$, where $c_0$ is
some specific constant (see {\cite{FPa}}, {\cite{PP}}). As a consequence we
know that there is always partial wetting, i.e., that inequality (\ref{pw}) is
satisfied, if the temperature is sufficiently low. Notice that the surface
tension $\tau^{+ -} ( \beta )$ is also analytic at low temperatures, and that
$\tau^{\text{\textrm{w}} -} - \tau^{\text{\textrm{w}} +} = 2 K$ and $\tau^{+
-} = 2 J$ when the temperature is zero.

Let us also mention that, for models including the one under consideration
that, the microscopic validity of the Winterbottom construction, has been
established, within a canonical ensemble when the size of the drop tends to
infinity, in ref. {\cite{BIV}}.

\tmfloat{h}{big}{figure}{\epsfig{file=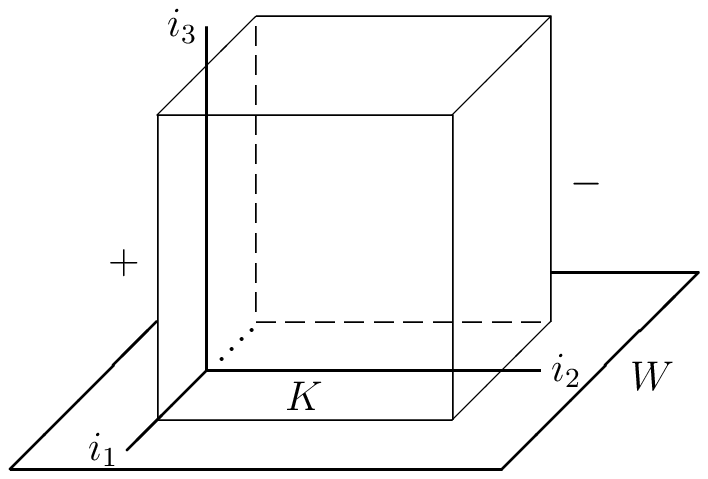}}{\small The box $V$ and the wall $W$ with
the $( + - )$ boundary condition.}

We next introduce, using the above framework, the definition of the line
tension, see Fig. 2. Consider the box
\begin{equation}
  V = \{ i \in {\mathbb{L}} : - L_1 + 1 \le i_1 \le L_1, 1 \le i_2 \le
  L_2, 1 \le i_3 \le L_3 \} \label{volume}
\end{equation}
and the boundary condition $\bar{\tmmathbf{\sigma}} = \{ \sigma_i =
\bar{\sigma}_i, i \in {\mathbb{L}} \}$. The associated partition
function is
\begin{equation}
  Z^{\bar{\tmmathbf{\sigma}}} ( V, \beta ) = \sum_{\tmmathbf{\sigma}_V} \exp
  \left( - \beta H_V (\tmmathbf{\sigma}_V \mid \bar{\tmmathbf{\sigma}} )
  \right)
\end{equation}
where the sum runs over all configurations, $\tmmathbf{\sigma}_V = \{
\sigma_i, i \in V \}$, inside $V$. The hamiltonian, according to the above
description of the model, is
\begin{equation}
  H_V (\tmmathbf{\sigma}_V \mid \bar{\sigma} ) = - J
  \sum_{\tmscript{\begin{array}{c}
    \langle i, j \rangle\\
    i, j \in V
  \end{array}}} ( \sigma_i \sigma_j - 1 ) - J \sum_{\tmscript{\begin{array}{c}
    \langle i, j \rangle\\
    i \in V, j \in \mathbb{L} \setminus V
  \end{array}}} ( \sigma_i \bar{\sigma}_j - 1 ) - K \sum_{i \in V, i_3 = 1}
  \sigma_i \label{ham}
\end{equation}
where the fist and second sum are over the nearest neighbor pairs $\langle i,
j \rangle$. We shall, in particular, consider the $+$ and $-$ boundary
conditions where $\bar{\sigma}_i = + 1$ and $\bar{\sigma}_i = - 1$, for all $i
\in {\mathbb{L}}$. And also the mixed $+ -$ boundary condition where
$\bar{\sigma}_i = + 1$ for $i_1 \ge 1$ and $\bar{\sigma}_i = - 1$ for $i_1 \le
0$.

We consider also the usual Ising model on the lattice
${\mathbb{Z}}^3$, inside the box
\begin{equation}
  V' = \{ i \in {\mathbb{Z}^3} : - L_1 + 1 \le i_1 \le L_1, 1 \le i_2
  \le L_2, - L_3 + 1 \le i_3 \le L_3 \},
\end{equation}
the hamiltonian $\tilde{H}_{V'} (\tmmathbf{\sigma}_{V'} \mid \bar{\sigma} )$
being given by an expression which contains only the first and second terms of
formula (\ref{ham}), replacing $\mathbb{L}$ by $\mathbb{Z}^3$ in the second
sum. We shall denote by $\tilde{Z}^{\bar{\tmmathbf{\sigma}}} ( V', \beta )$
the associated partition functions, and consider also the $+$, $-$ and $+ -$
boundary conditions, defined as before, for all $i \in \mathbb{Z}^3$.

With the above notations, the definition of the line tension is as follows
\begin{equation}
  \lambda ( \beta ) = \lim_{L_2 \to \infty} \lim_{L_3 \to \infty} \lim_{L_1
  \to \infty} - \frac{1}{\beta L_2} \ln \frac{Q ( V, \beta )}{\left( \tilde{Q}
  ( V', \beta ) \right)^{1 / 2}} \label{line}
\end{equation}
where
\begin{eqnarray}
  Q ( V, \beta ) & = & Z^{+ -} ( V, \beta ) / \left( Z^+ ( V, \beta ) Z^- ( V,
  \beta ) \right)^{1 / 2}  \label{Q1}\\
  \tilde{Q} ( V', \beta ) & = & \tilde{Z}^{+ -} ( V', \beta ) / \tilde{Z}^+ (
  V', \beta )  \label{Q2}
\end{eqnarray}
In order to explain this definition we first recall that $\ln \tilde{Q} ( V',
\beta )$ represents the free energy of the $+ -$ interface, see {\cite{LP}}.
The mixed boundary condition $+ -$ forces the system to produce, inside the
volume $V'$, a separation between the $( + )$ phase at the left side and the
$( - )$ phase at the right side. In expression (\ref{Q2}) the volume
contributions, proportional to the free energy of the $( + )$ and $( - )$
phases, as well as the boundary effects, cancel, and only the contributions to
the free energy due to the interface are left. We recall also the definition
of the surface tension between the $( - )$ phase and the substrate, see
{\cite{FPa}},
\begin{equation}
  \tau^{\text{\textrm{w}} -} ( \beta ) = \lim_{L_1, L_2 \to \infty} \lim_{L_3
  \to \infty} - \frac{1}{\beta L_1 L_2} \ln \frac{Z^- ( V, \beta )}{\left(
  \tilde{Z}^- ( V', \beta ) \right)^{1 / 2}} \label{W}
\end{equation}
The boundary condition forces the volume $V$ to be occupied by the $( - )$
phase, and thus the logarithm in equation (\ref{W}) represents the
contribution to the free energy due to the interface between the $( - )$ phase
and the wall. The surface tension between the $( + )$ phase and the substrate
is defined analogously.

Similar arguments show then that $\ln Q ( V, \beta )$, defined by (\ref{Q1}),
corresponds to the contribution to the free energy due to the $+ -$ interface,
inside the box $V$, together with the contribution due to the line of contact
of this interface with the wall. It follows, therefore, that definition
(\ref{line}) represents, as stated, the free energy per unit length of this
line.

There is no energy associated to the line, so the line tension given by
equation (\ref{line}), equals $0$ at zero temperature. When the temperature is
sufficiently low, the line tension can be rigorously studied, using cluster
expansion techniques, and this is essentially the content of the present work.
The main results are summarized in Theorem \ref{T1} below. Let us mention that
the cluster expansion associated to a rigid $+ -$ interface is needed in this
study and that, for this reason, we can only consider the line tension when
the line is parallel to a lattice axis. It is expected, however, that the line
tension exists for any orientation of the line and has a similar behavior.
Notice also that we know, see {\cite{Mir}}, that the equilibrium shape of the
drop, associated to the Ising model at low temperatures, has facets parallel
to the coordinate planes, as shown in Fig. 1, and so the portions of the line
parallel to an axis have a positive length.

\begin{theorem}
  \label{T1}For $J \ge |K|$ and if the temperature is low enough, i.e., if
  $\beta ( J - |K| ) \ge c_0$, where $c_0$ is a given constant, then the line
  tension, $\lambda ( \beta )$, exists and is strictly negative as soon as the
  temperature is different from zero. Moreover, $\lambda ( \beta )$ is an
  analytic function for which an explicit convergent series expansion can be
  found, whose leading term is
  \begin{equation}
    - \frac{2}{\beta} e^{- 6 \beta J} \cosh ( 2 \beta K )
  \end{equation}
\end{theorem}

The proof is given in section 2.

\begin{remark}
  The series is given by expressions (\ref{lt}), (\ref{scl}), (\ref{sagg}) and
  (\ref{saggr}), below. An easy bound is $c_0 = 2 \kappa_{cl} + \ln ( 5 \nu )
  + \ln ( e - 1 )$, where $\kappa_{cl} = a_0 + \ln \left( ( 1 + a_0 ) / a_0
  \right) \sim 1.58$ is the cluster constant, $a_0 = ( \sqrt{5} - 1 ) / 2$,
  and $\nu = ( 12 )^2$.
\end{remark}

\begin{remark}
  The method of the proof of the theorem, can be used to show, by adding some
  natural ingredients, that the interface associated to the $+ -$ boundary
  condition, and hence the contact line, is rigid at low temperature.
\end{remark}

\section{Proof of theorem}

\subsection{Contours}

We begin with a contour representation of the partition functions $Z^+ ( V,
\beta ), Z^- ( V, \beta )$. A natural definition is to consider contours as
boundaries of regions where the considered configuration differs from the
corresponding ground state configuration: the $+$ configuration with the $+$
boundary condition and the $-$ configuration with the $-$ boundary condition.

For $Z^+ ( V, \beta )$ we have a standard representation introducing for any
configuration $\tmmathbf{\sigma}$ (such that $\sigma_i = + 1$ for $i \in
\mathbb{L} \setminus V$) the contours as connected components of the set $B^+
( \tmmathbf{\sigma} )$ of all plaquettes of the dual lattice that separate two
neighbouring sites $i, j \in \mathbb{L} \cup W$ with $\sigma_i \neq
\sigma_j$.

For any contour $\gamma$ we introduce the weight factor
\begin{equation}
  z^+ ( \gamma ) = e^{- 2 \beta J| \gamma^{\tmop{bk}} | - 2 \beta K| \gamma^W
  |} \label{poids+}
\end{equation}
Here $\gamma^W$ is the set of plaquettes of $\gamma$ that separate a site of
the wall from a site of the first layer $\mathbb{L}_1 = \{ ( x, y, z ) : z =
1 \}$, $\gamma^{\tmop{bk}} = \gamma \setminus \gamma^W$, $| \gamma^{\tmop{bk}}
|$ and $| \gamma^W |$ denote repectively the number of plaquettes of
$\gamma^{\tmop{bk}}$ and $\gamma^W$. In terms of the weight factor $z^+ (
\gamma )$, one has
\begin{equation}
  \label{fp+} Z^+ ( V, \beta ) = e^{\beta K|W ( V ) |} \sum_{\{ \gamma_1,
  \ldots, \gamma_n \}_{\op{\tmop{comp}}} \subset V} \prod_{i = 1}^n z^+ (
  \gamma_i )
\end{equation}
where $W ( V )$ is the set of sites of the wall that have a nearest neighbor
in $V$ and $\{ \gamma_1, \ldots, \gamma_n \}_{\op{\tmop{comp}}}$ is a
collection of compatible (mutually disjoint) contours in $V$: this means more
precisely, that the considered contours consits of set of plaquettes dual to
n.n. pairs containing a site in $V$.

To get a similar expression for $Z^- ( V, \beta )$, we only have to be
carefull with the definition of {\tmem{contours touching the substrate}}:
those are the contours that contain a plaquette dual of a n.n.\
pair with one site on the substrate $W$ and this means that they do
not intersect the plane $i_3 = 1 / 2$. Namely, for configurations
$\tmmathbf{\sigma}$ such that $\sigma_i = + 1$ for $i \in W$ and $\sigma_i = -
1$ for $i \in \mathbb{L} \setminus V$, we introduce contours as connected
components of the set $B^- ( \tmmathbf{\sigma} )$ of all plaquettes separating
nearest neighbor sites $i, j \in V$ such that $\sigma_i \neq \sigma_j$ or
nearest neighbor sites $i \in V$, $j \in W$ for which $\sigma_i = \sigma_j ( =
+ 1 )$. Introducing now the weight $z^- ( \gamma )$ as
\begin{equation}
  \label{poids-} z^- ( \gamma ) = e^{- 2 \beta J| \gamma^{\tmop{bk}} | + 2
  \beta K| \gamma^W |}
\end{equation}
we get
\begin{equation}
  \label{fp-} Z^- ( V, \beta ) = e^{- \beta K|W ( V ) |} \sum_{\{ \gamma_1,
  \ldots, \gamma_n \}_{\op{\tmop{comp}}} \subset V} \prod_{i = 1}^n z^- (
  \gamma_i )
\end{equation}
Notice that the set of contours in both situations exactly coincide (even
though the weights do not) and the sums in (\ref{fp+}) and (\ref{fp-}) are
over exactly the same collections of contours. Notice also that the weights
(\ref{poids+}) and (\ref{poids-}) differ only if the contour $\gamma$ touches
the substrate.

For the partition function $\tilde{Z}^+ ( V' \beta )$ of the Ising model we
have the standard expansion
\begin{equation}
  \tilde{Z}^+ ( V', \beta ) = \sum_{\{ \gamma_1, \ldots, \gamma_n
  \}_{\op{\tmop{comp}}} \subset V'} \prod_{i = 1}^n z ( \gamma_i )
\end{equation}
where
\begin{equation}
  z ( \gamma ) = e^{- 2 \beta J| \gamma |} \label{poids}
\end{equation}
and the sum is over compatible families of contours in the box $V'$: the
contours here are connected components of the set $B^+ ( \tmmathbf{\sigma} )$
of all plaquettes of the dual lattice that separate two neighbouring sites $i,
j \in \mathbb{Z}^3$ with $\sigma_i \neq \sigma_j$, for configurations
$\tmmathbf{\sigma}$ such that $\sigma_i = + 1$ for  $i \in \mathbb{Z}^3
\setminus V'$.

To be able to control, in terms of convergent cluster expansions, the
logarithm of the above partition functions, we need good decaying behavior of
the activities of contours with respect to their area. It is easy to realize
from geometrical observations ($| \gamma^{\tmop{bk}} | > | \gamma^W |$) that
\begin{equation}
  z^{\pm} ( \gamma ) \leqslant e^{- \beta ( J - |K| ) | \gamma |}
  \label{borne}
\end{equation}
while for contours not touching the substrate one has obviously
\begin{equation}
  z^{\pm} ( \gamma ) = z ( \gamma ) = e^{- 2 \beta J| \gamma |}
\end{equation}

\subsection{Multi-indexes and clusters}

We now introduce {\tmem{multi--indexes}} in order to write the logarithm of
the partition functions $Z^+ ( V, \beta )$ and $Z^- ( V, \beta )$ as a sum
over these multi-indexes {\cite{GMM}}. A multi-index $X$ is a function from
the set of contours (in $V$) into the set of nonnegative integers. We let
$\op{\tmop{supp}} X = \cup_{\gamma : X ( \gamma ) \geq 1} \gamma$ denotes the
support of the multi--index X and $|X| = \sum_{\gamma : X ( \gamma ) \geq 1} X
( \gamma ) | \gamma |$ denotes its area. For the activities $z^+$and $z^-$, we
define the truncated functions

\begin{equation}
  \Phi^{^{^{}} \pm} ( X ) = \frac{a ( X )}{\prod_{\gamma} X ( \gamma ) !}
  \prod_{\gamma} z^{\pm} ( \gamma )^{X ( \gamma )}
\end{equation}
where the factor $a ( X )$ is a combinatoric factor defined in terms of the
connectivity properties of the graph $G ( X )$ with vertices corresponding to
$\gamma \in \op{\tmop{supp}} X$ (there are $X ( \gamma )$ vertices for each
$\gamma \in \tmop{supp} X$) that are connected by an edge whenever the
corresponding contours are incompatibles. Namely, $a ( X ) = 0$ and hence
$\widetilde{\Phi_{}}^{^{^{}} \pm}_K ( X ) = 0$ unless $G ( X )$ is a connected
graph or equivalently $\tmop{supp} X$ is a connected set, and
\begin{equation}
  a ( X ) = \sum_{G \subset G ( X )} ( - 1 )^{|e ( G ) |} \label{multi}
\end{equation}
Here the sum goes over connected subgraphs $G$ whose vertices coincide with
the vertices of $G ( X )$ and $|e ( G ) |$ is the number of edges of the graph
$G$. The connected multi--indexes will be call {\tmem{clusters}}. Whenever $X$
contains only one contour $\gamma$ (i.e. $X ( \gamma ) = 1$ and $X ( \gamma' )
= 0$ for all others contours), then $a ( X ) = 1$, implying $\Phi^{^{^{}} \pm}
( X ) = z^{\pm} ( \gamma )$  in such a case.

We will say that a multi--index or a cluster $X$ touches the substrate if
there exists a contour in the support of $X$ touching the substrate. Notice
that for multi--indexes or clusters $X$ supported by contours not touching the
substrate (we will say that $X$ do not touch the substrate), one has
$\Phi^{^{^{}} +}_K ( X ) = \Phi^{^{^{}} -}_K ( X )$.

A consequence of the previous definitions is that the sums entering in the
expressions of the partition functions $\text{$Z^+ ( V, \beta )$}$ and
$\text{$Z^- ( V, \beta )$}$ can be exponentiated as follows
\begin{equation}
  \sum_{\{ \gamma_1, \ldots, \gamma_n \}_{\op{\tmop{comp}}} \subset V}
  \prod_{i = 1}^n z^{\pm} ( \gamma_i ) = \exp \{ \sum_{X \subset V}
  \Phi^{^{^{}} \pm} ( X ) \} \label{clusterfp}
\end{equation}
where the sum runs over non--empty clusters (supported by contours) in the box
$V$.

In addition, one has the following convergence properties.

\begin{lemma}
  \label{l:cluster} Assume that $\beta ( J - |K| ) \geqslant \log \nu +
  \kappa_{\tmop{cl}}$, then
  \begin{equation}
    \sum_{X : X ( \gamma ) \geq 1} | \Phi^{^{^{}} \pm} ( X ) | \leqslant e^{-
    [ \beta ( J \pm K ) - a_0 ] | \gamma |} \label{bornephi}
  \end{equation}
  and the series $\sum_{X : \tmop{supp} X \ni i} | \Phi^{^{^{}} \pm} ( X ) |$
  is absolutely convergent.
\end{lemma}

\begin{proof}
  We first notice that the numbers of contours $\gamma$ of area $| \gamma | =
  n$ containing a given vertex is less than $\nu^n$.
  
  Under the condition
  \begin{equation}
    z^{\pm} ( \gamma ) \leqslant ( e^{\mu^{\pm} ( \gamma )} - 1 ) \exp [ -
    \sum_{\gamma' \nsim \gamma} \mu^{\pm} ( \gamma ) ] \label{condconv}
  \end{equation}
  where $\mu$ is a positive function and the sum is over contours $\gamma'$
  incompatible with the contour $\gamma$ (the relation denoted $\gamma' \nsim
  \gamma$ means that $\gamma'$ does not intersect $\gamma$), we know from Ref.
  {\cite{M}}, that the truncated functions $\Phi^{^{^{}} \pm} ( X )$ satisfies
  the estimate
  \begin{equation}
    \sum_{X : X ( \gamma ) \geq 1} | \Phi^{^{^{}} \pm}_K ( X ) | \leqslant
    \mu^{\pm} ( \gamma ) \label{borne2}
  \end{equation}
  We choose $\mu^{\pm} ( \gamma ) = e^{- [ \beta ( J \pm K| ) - a ] | \gamma
  |}$ to get by taking into account the above remark on the entropy of
  contours and that the minimal area of contours is $6$, that
  \begin{equation}
    \sum_{\gamma' \nsim \gamma} \mu^{\pm} ( \gamma ) \leqslant 2| \gamma |
    \sum_{n \equallim 6}^{\infty} \nu^n e^{- [ \beta ( J \pm K ) - a ] n}
    \leqslant \frac{1}{e^{[ \beta ( J \pm K ) - a - \log \nu ]} - 1} | \gamma
    | \label{borne3}
  \end{equation}
  provided $2 \nu^5 e^{- 5 [ \beta ( J \pm K ) - a ]} \leqslant 1$. The factor
  $2$ stems from the fact that a contour of area $| \gamma |$ contains at most
  $2| \gamma |$ vertices. Since $\mu^{\pm} ( \gamma ) \leqslant e^{\mu^{\pm} (
  \gamma )} - 1$, the bound (\ref{borne}) on the activities of contours gives
  that the convergence condition (\ref{condconv}) will be satisfied whenever
  \begin{equation}
    \beta ( J \pm K ) \geqslant \log \nu + a + \log \frac{1 + a}{a}
  \end{equation}
  The choice $a = a_0$, that minimizes the function $a + \log \frac{1 +
  a}{a}$, and for which $2 \nu^5 e^{- 5 [ \beta ( J - |K| ) - a ]} \leqslant
  1$ provides the condition given in the lemma. 
\end{proof}

Note that Lemma \ref{l:cluster} gives the bound:
\begin{equation}
  | \Phi^{^{^{}} \pm} ( X ) | \leqslant_{} e^{- [ \beta ( J \pm K ) - a_1 ]
  |X|} \label{bornephi3}
\end{equation}
with $a_1 = \kappa_{\tmop{cl}} + \log \nu$.

To exponentiate the partition function $\tilde{Z} ( V', \beta )$, we
introduce the truncated function associated with the activity $z$:
\[ \Phi ( X ) = \frac{a ( X )}{\prod_{\gamma} X ( \gamma ) !} \prod_{\gamma} z
   ( \gamma )^{X ( \gamma )} \]
where the multi--indexes are define here as functions from the set of contours
in $V'$ into the set of non negative integers. One has
\begin{equation}
  \tilde{Z} ( V', \beta ) = \exp \{ \sum_{X \subset V'} \Phi ( X ) \}
  \label{clusterfpp}
\end{equation}
and
\begin{equation}
  | \Phi ( X ) | \leqslant_{} e^{- ( 2 \beta J - a_1 ) |X|}
\end{equation}

\subsection{Interfaces\label{interfaces}}

We now turn to the partition function $Z^{+ -} ( V, \beta )$ that we will
expand in terms of interfaces. Let $W^+$ be the set of the sites $i = ( i_1,
i_2, 0 )$ of the wall with $i_1 \geqslant 1$ and let $W^- = W \setminus W^+$
denotes its complement. For a configuration $\tmmathbf{\sigma}$ that coincides
with the $+ -$ boundary conditon outside the box $V$, consider the set $B^{+
-} ( \tmmathbf{\sigma} )$ of all plaquettes separating, neighboring sites $i$
$j \in \mathbb{L}$ with $\sigma_i \neq \sigma_j$, neighboring sites $i \in V,
j \in W^+$ with $\text{$\sigma_i \neq \sigma_j$}$, and neighboring sites $i
\in V, j \in W^-$ with $\text{$\sigma_i = \sigma_j ( = + 1 )$}$. We decompose
this set into maximal connected components. There is exactly one component
which is infinite. We call this component $I$ the interface. The possible
interfaces are the sets $I \in \mathcal{I}$ for which there exists a
configuration $\tmmathbf{\sigma}$ such that $I = B^{+ -} (\tmmathbf{\sigma})$.

Notice that this set is the same as the set of interfaces of the Ising model
in a box $V$ included in the lattice $\mathbb{Z}^3$, with the $+ -$ boundary
conditions, considered by Dobrushin {\cite{D1}}. We will consider this set at
the end of the subsection for the box $V'$.

We let $I^{\tmop{bk}}$ denotes the set of plaquettes of $I$ dual to pairs $i
\in \mathbb{L}, j \in V \tmop{or} j \in \mathbb{L}$, $I^{W^+}$ denotes the
set of plaquettes of $I$ dual to pairs $i \in V, j \in W^+$ and $I^{W^-}$
denotes the set of plaquettes dual to pairs $i \in V, j \in W^- $.

The interface $I$ divides the set $V$ into two subsets $V_+ = V_+ ( I )$ and
$V_- = V_- ( I )$:  $V_+$ (respectively $V_-$) is the part of $V$ of the $+$
sites ($\sigma_i = + 1 )$  (respectively of the $-$sites ($\sigma_i = - 1$))
of the configuration $\tmmathbf{\sigma}$ such that $I = B^{+ -}
(\tmmathbf{\sigma})$.

With these definitions, we get the following expansion:
\begin{eqnarray*}
  Z^{+ -} ( V, \beta ) & = & \sum_I e^{- 2 \beta J|I^{\tmop{bk}} |} e^{- 2
  \beta K| I^{W^+} |} e^{+ \beta K|W^+ \cap W ( V ) |}
  \sum_{\tmscript{\begin{array}{l}
    \{ \gamma_1, \ldots, \gamma_n \}_{\op{\tmop{comp}}} \subset V_+\\
    \{ \gamma_1, \ldots, \gamma_n \}_{\op{\tmop{comp}}} \sim I
  \end{array}}} \prod_{i = 1}^n z^+ ( \gamma_i )\\
  &  & \times e^{2 \beta K| I^{W^-} |} e^{- \beta K|W^- \cap W ( V ) |}
  \sum_{\tmscript{\begin{array}{l}
    \{ \gamma_1, \ldots, \gamma_n \}_{\op{\tmop{comp}}} \subset V_-\\
    \{ \gamma_1, \ldots, \gamma_n \}_{\op{\tmop{comp}}} \sim I
  \end{array}}} \prod_{i = 1}^n z^- ( \gamma_i )
\end{eqnarray*}
where the two last sums are over collections of contours compatible with the
interface: the compatibility relation is denoted $\sim$ and means that no
contour of the considered collections intersects the interface.

By taking into account (\ref{clusterfp}), we obtain:
\begin{eqnarray*}
  Z^{+ -} ( V, \beta ) & = & \sum_I \exp \{ - 2 \beta \text{$J|I^{\tmop{bk}}$}
  | - 2 \beta K|I^{W^+} | + 2 \beta K|I^{W^-} | \}\\
  &  & \times \exp \{ \sum_{\tmscript{\begin{array}{l}
    X \subset V_+\\
    X \sim I
  \end{array}}} \Phi^{ +} ( X ) + \sum_{\tmscript{\begin{array}{l}
    X \subset V_-\\
    X \sim I
  \end{array}}}\Phi^{ -} ( X ) \}
\end{eqnarray*}
where the two last sums are over clusters compatible with the interface. We
then have, for the ratio (\ref{Q1}), using (\ref{fp+}), (\ref{fp-}), and
(\ref{clusterfp}):
\begin{eqnarray}
  Q ( V, \beta ) & = & \sum_I \exp \{ - 2 \beta \text{$J|I^{\tmop{bk}}$} | - 2
  \beta K|I^{W^+} | + 2 \beta K|I^{W^-} | \}  \label{interface}\\
  & \times & \exp \{ \sum_{\tmscript{\begin{array}{l}
    X \subset V_+\\
    X \sim I
  \end{array}}} \Phi^{+} ( X ) + \sum_{\tmscript{\begin{array}{l}
    X \subset V_-\\
    X \sim I
  \end{array}}} \Phi^{-} ( X ) - \frac{1}{2} \sum_{X \subset V_{}}
  \Phi^{ +} ( X ) - \frac{1}{2} \sum_{X \subset V} \Phi^{ -} (
  X ) \} \nonumber
\end{eqnarray}
We put
\begin{equation}
  A_{} ( I, V ) = \sum_{\tmscript{\begin{array}{l}
    X \subset V_+\\
    X \sim I_{}
  \end{array}}} \Phi^{^{^{}} +} ( X ) + \sum_{\tmscript{\begin{array}{l}
    X \subset V_-\\
    X \sim I
  \end{array}}} \Phi^{^{^{}} -} ( X ) - \frac{1}{2} \sum_{X \subset V_{}}
  \Phi^{^{^{}} +} ( X ) - \frac{1}{2} \sum_{X \subset V_{}} \Phi^{^{^{}} -} (
  X )
\end{equation}
and we will decompose the first sum into sum over clusters not touching the
substrate, we write $X \sim W^+$ and sum over clusters touching the substrate,
we write $X \nsim W^+$. We make analogous decompostions for the three other
sums. Taking into account that for clusters $X$ not touching the substrate,
one has $\Phi^{^{^{}} +} ( X ) = \Phi^{^{^{}} -} ( X ) = \Phi^{^{^{}}}_{} ( X
)$, we get:
\begin{eqnarray}
  A ( I, V ) & = & \sum_{\tmscript{\begin{array}{l}
    X \subset V_+\\
    X \sim I\\
    X \sim W^+
  \end{array}}} \Phi ( X ) + \sum_{\tmscript{\begin{array}{l}
    X \subset V_+\\
    X \sim I\\
    X \nsim W^+
  \end{array}}} \Phi^+_K ( X ) + \sum_{\tmscript{\begin{array}{l}
    X \subset V_-\\
    X \sim I\\
    X \sim W^-
  \end{array}}} \Phi ( X ) + \sum_{\tmscript{\begin{array}{l}
    X \subset V_-\\
    X \sim I\\
    X \nsim W^-
  \end{array}}} \Phi^- ( X ) \nonumber\\
  &  & - \sum_{\tmscript{\begin{array}{l}
    X \subset V^{}\\
    X \sim W
  \end{array}}} \Phi ( X ) - \frac{1}{2} \sum_{\tmscript{\begin{array}{l}
    X \subset V^{}\\
    X \nsim W
  \end{array}}} \Phi^+ ( X ) - \frac{1}{2} \sum_{\tmscript{\begin{array}{l}
    X \subset V\\
    X \nsim W
  \end{array}}} \Phi^- ( X ) \nonumber\\
  &  &  \nonumber\\
  & = & - \sum_{\tmscript{\begin{array}{l}
    X \subset V\\
    X \nsim I\\
    X \sim W_{}
  \end{array}}} \Phi ( X ) + \sum_{\tmscript{\begin{array}{l}
    X \subset V_+\\
    X \sim I\\
    X \nsim W^+
  \end{array}}} \Phi^+ ( X ) + \sum_{\tmscript{\begin{array}{l}
    X \subset V_-\\
    X \sim I\\
    X \nsim W^-
  \end{array}}} \Phi^- ( X ) \nonumber\\
  &  & - \frac{1}{2} \sum_{\tmscript{\begin{array}{l}
    X \subset V\\
    X \nsim W
  \end{array}}} \Phi^+ ( X ) - \frac{1}{2} \sum_{\tmscript{\begin{array}{l}
    X \subset V\\
    X \nsim W
  \end{array}}} \Phi^- ( X )  \label{adei}
\end{eqnarray}
where the first sum in the last term is over clusters inside $V$ incompatible
with the interface. We then decompose the last two terms as follows:
\begin{eqnarray*}
  \sum_{\tmscript{\begin{array}{l}
    X \subset V^{}\\
    X \nsim W
  \end{array}}} \Phi^+ ( X ) & = & \sum_{\tmscript{\begin{array}{l}
    X \subset V\\
    X \nsim W^+\\
    X \nsim W^-
  \end{array}}} \Phi^+ ( X ) + \sum_{\tmscript{\begin{array}{l}
    X \subset V\\
    X \nsim W^+\\
    X \sim W^-
  \end{array}}} \Phi^+ ( X ) + \sum_{\tmscript{\begin{array}{l}
    X \subset V\\
    X \nsim W^-\\
    X \sim W^+
  \end{array}}} \Phi^+ ( X )\\
  & = & - \sum_{\tmscript{\begin{array}{l}
    X \subset V\\
    X \nsim W^+\\
    X \nsim W^-
  \end{array}}} \Phi^+ ( X ) + 2 \sum_{\tmscript{\begin{array}{l}
    X \subset V\\
    X \nsim W^+\\
    X \nsim W^-
  \end{array}}} \Phi^+ ( X ) + 2 \sum_{\tmscript{\begin{array}{l}
    X \subset V\\
    X \nsim W^+\\
    X \sim W^-
  \end{array}}} \Phi^+ ( X )\\
  & = & - \sum_{\tmscript{\begin{array}{l}
    X \subset V\\
    X \nsim W^+\\
    X \nsim W^-
  \end{array}}} \Phi^+ ( X ) + 2 \sum_{\tmscript{\begin{array}{l}
    X \subset V\\
    X \nsim W^+
  \end{array}}} \Phi^+ ( X )\\
  \sum_{\tmscript{\begin{array}{l}
    X \subset V^{}\\
    X \nsim W
  \end{array}}} \Phi^- ( X ) & = & - \sum_{\tmscript{\begin{array}{l}
    X \subset V\\
    X \nsim W^+\\
    X \nsim W^-
  \end{array}}} \Phi^- ( X ) + 2 \sum_{\tmscript{\begin{array}{l}
    X \subset V\\
    X \nsim W^-
  \end{array}}} \Phi^- ( X )
\end{eqnarray*}
(using that $\sum_{\tmscript{\begin{array}{l}
  X \subset V\\
  X \nsim W^-\\
  X \sim W^+
\end{array}}} \Phi^{\pm} ( X ) = \sum_{\tmscript{\begin{array}{l}
  X \subset V\\
  X \sim W^-\\
  X \nsim W^+
\end{array}}} \Phi^{\pm} ( X )$). Inserting the two previous equalities in
(\ref{adei}) and using that
\[ \sum_{\tmscript{\begin{array}{l}
     X \subset V_{\pm}\\
     X \sim I_{}\\
     X \nsim W^{\pm}
   \end{array}}} \Phi^{\pm} ( X ) - \sum_{\tmscript{\begin{array}{l}
     X \subset V_{}\\
     X \nsim W^{\pm}
   \end{array}}} \Phi^{\pm} ( X ) = - \sum_{\tmscript{\begin{array}{l}
     X \subset V_{}\\
     X \nsim I_{}\\
     X \nsim W^{\pm}
   \end{array}}} \Phi^{\pm} ( X ) \]
gives
\[ A ( I, V ) = \ell ( V ) + B ( I, V ) \]
where
\begin{equation}
  \ell ( V ) = \frac{1}{2} \sum_{\tmscript{\begin{array}{l}
    X \subset V\\
    X \nsim W^+\\
    X \nsim W^-
  \end{array}}} [ \Phi^+ ( X ) + \Phi^- ( X ) ] \label{aaa}
\end{equation}
and
\begin{equation}
  B_{} ( I, V ) = - \sum_{\tmscript{\begin{array}{l}
    X \subset V^{}\\
    X \nsim I\\
    X \sim W_{}
  \end{array}}} \Phi ( X ) - \sum_{\tmscript{\begin{array}{l}
    X \subset V_{}\\
    X \nsim I_{}\\
    X \nsim W^+
  \end{array}}} \Phi^+ ( X ) + \sum_{\tmscript{\begin{array}{l}
    X \subset V_{}\\
    X \nsim I\\
    X \nsim W^-
  \end{array}}} \Phi^- ( X ) \label{bbb}
\end{equation}

In order to analyse the interface, we consider the system in the infinite
cylinder
\[ \bar{V} = \lim_{L_1 \rightarrow \infty} V = \{ i = ( i_1, i_2, i_3 ) \in
   \mathbb{L} : i_1 \in \mathbb{Z}, 0 \leq i_2 \leq L_2, 1 \leq i_3 \leq L_3
   \} \]
The absolute convergence of the series of truncated functions implies the
existence of the limit, $\lim_{L_1 \rightarrow \infty} Q ( V, \beta )$. We
denote this limit by $Q ( \bar{V}, \beta )$. One has
\begin{equation}
  \text{$Q ( \bar{V}, \beta )$} = e^{\ell ( \bar{V} )} \sum_I \exp \{ - 2
  \beta \text{$J|I^{\tmop{bk}}$} | - 2 \beta K|I^{W^+} | + 2 \beta K|I^{W^-} |
  + B ( I, \bar{V} ) \} \label{rapportk}
\end{equation}
where the sum is now over interfaces in $\bar{V}$, $\ell ( \bar{V} )$ and $B (
I, \bar{V} )$ are respectively defined by (\ref{aaa}) and (\ref{bbb}) with $V$
replaced by $\bar{V}$.

We next consider the partition function $\tilde{Z}^{+ -} ( V', \beta )$, and
for a configuration $\tmmathbf{\sigma}$ that coincides with the $+ -$ boundary
conditon outside the box $V'$, we consider the set $B^{+ -} (
\tmmathbf{\sigma} )$ of all plaquettes separating, neighboring sites $i$ $j
\in \mathbb{Z}^3$ with $\sigma_i \neq \sigma_j$. Again, there is exactly one
component which is infinite. We denote this componant $\tilde{I}$. We then get
the following expansion (see {\cite{D1}} or {\cite{Mir}})
\begin{equation}
  \text{$\tilde{Z}^{+ -} ( V', \beta )$} = \sum_{\tilde{I}} \exp \{ - 2 \beta
  \text{$J| \tilde{I}$} | + \sum_{X \subset V'} \Phi ( X ) \}
\end{equation}
over interfaces, which leads to
\begin{equation}
  \text{$\tilde{Q} ( V', \beta )$} = \sum_{\tilde{I}} \exp \{ - 2 \beta
  \text{$J| \tilde{I}$} | - \sum_{X \nsim \tilde{I}, X \subset V'} \Phi ( X )
  \}
\end{equation}
where the sum in the exponant is over clusters in $V'$ incompatible with the
interface $\tilde{I}$. We introduce the infinite box
\[ \bar{V}' = \lim_{L_1 \rightarrow \infty} V' = \{ i = ( i_1, i_2, i_3 ) \in
   \mathbb{Z}^3 : i_1 \in \mathbb{Z}, 0 \leq i_2 \leq L_2, 1 \leq i_3 \leq
   L_3 \} \]
Then, denoting by $\tilde{Q} ( \bar{V}', \beta )$ the limit $\lim_{L_1
\rightarrow \infty} \tilde{Q} ( V', \beta )$, one has:
\begin{equation}
  \text{$\tilde{Q} ( \bar{V}', \beta )$} = \sum_{\tilde{I}} \exp \{ - 2 \beta
  \text{$J| \tilde{I}$} | - \sum_{X \nsim \tilde{I}, X \subset \bar{V}'} \Phi
  ( X ) \} \label{rapport}
\end{equation}

\subsection{Walls\label{walls}}

In this subsection, following Ref. {\cite{D1}}, we will described the
interfaces $I$ and $\tilde{I}$ appearing in equations (\ref{rapportk}) and
(\ref{rapport}) in terms of exitations called walls. We begin with the
interfaces $I$.

Let $P$ be the horizontal plane $i_1 = 1 / 2$ and $\pi ( \cdot )$ the
orthogonal projection on this plane. The projection $\pi ( p )$ of a plaquette
is either a plaquette or an edge. There are two types of plaquettes in an
interface $I$: the {\tmem{ceiling plaquettes}}, which are the plaquettes $p$
parallel to the plane $P$ and such that there is no other plaquette $p'$ such
that $\pi ( p ) = \pi ( p' )$, and the {\tmem{wall plaquettes}}, which are all
the other plaquettes in $I$. The set of wall plaquettes is denoted
$\mathcal{W}( I )$.

A (connected) set $w$ of wall plaquettes is called a {\tmem{standard wall}}
(or {\tmem{wall in standard position}}) if there exists an interface $I$ such
that $w =\mathcal{W}( I )$. A family of standard walls is {\tmem{admissible}}
if the projections on the plane $P$ of these walls are pairwise disjoint. It
will be seen that the interfaces can equivalently be described by the
admissible families of standard walls.

We observe that any interface $I$ decomposes into {\tmem{walls}}, which are
the subsets of $\mathcal{W}( I )$ which are projected into the maximally
connected components of the projection $\pi (\mathcal{W}( I ) )$, and
ceilings, or connected sets of ceiling plaquettes. Given a wall $w$, consider
the set $C$ of plaquettes on the plane $P$ which do not belong to $\pi ( p )$,
and decompose this set into connected components. To each component there
corresponds one ceiling adjacent to $w$ which projects into this component.
The ceiling which projects into the (unique) infinite component of $C$ is
called the {\tmem{{\tmem{base{\tmem{}}}}}} of $w$. Since the base of a
standard wall lies on $P$, one can associate with any wall $w$ the standard
wall which is just the translate of $w$ (with respect to the $i_1$--axis) with
base on $P$. In this way, one associates with every interface a family of
standard walls having disjoint projections on $P$ (i.e. those we have called
admissible). The converse is also true: for any admissible family $\{ w_1,
\ldots, w_n \}$ of standard walls, one can reconstruct in a unique way the
interface. This interface will be denoted $I ( w_1, \ldots, w_n \}$. 

Notice that a wall $w$ splits into three pieces, $w = \{ w^{\tmop{bk}},
w^{W^+}, w^{W^-} \}$: $w^{\tmop{bk}}$ is the set of plaquettes of $w$ dual to
n.n. pairs $i, j \in \mathbb{L}$, $w^{W^+}$ is the set of plaquettes of $w$
dual to n.n. pairs $i \in \mathbb{L}, j \in W^+$, $w^{W^-}$ being the set of
plaquettes of $w$ dual to n.n. pairs $i \in \mathbb{L}, j \in W^-$. We let:
\begin{equation}
  \text{$\rho ( w ) = \exp \{ - 2 \beta J|w^{\tmop{bk}} | + 2 \beta J| \pi (
  w^{\tmop{bk}} ) | - 2 \beta K|w^{W^+} | + 2 \beta K|w^{W^-} | \}$}
  \label{rrr}
\end{equation}
be the activity of $w$. Here $|w^{\tmop{bk}} |$, $| \pi ( w^{\tmop{bk}} ) |$,
and $|w^{W^{\pm}} |$ denote the number of plaquettes of the considered sets.
Note that the activities of walls depend on $K$ only for walls touching the
substrate ($w_{W^{\pm}} \neq \emptyset )$. We observe that:
\[ I^{\tmop{bk}} ( w_1, \ldots, w_n ) = L_2 L_3 + \sum_{i = 1}^n (
   |w_i^{\tmop{bk}} | - | \pi ( w_i^{\tmop{bk}} ) | ) \]
and that $I^{W^{\pm}} ( w_1, \ldots, w_n ) = \sum_{i = 1}^n |w_i^{W^{\pm}} |$.
Then, expression (\ref{rapportk}) becomes:
\begin{equation}
  \text{$Q ( \bar{V}, \beta )$} = e^{\ell ( \bar{V} ) - 2 \beta JL_1 L_3}
  \sum_{\{ w_1, \ldots, w_n \}_{\tmop{adm}} \subset \bar{V}} \prod_{i = 1}^n
  \rho ( w_i ) \exp \{ B ( I ( w_1, \ldots, w_n ), \bar{V} ) \} \label{qqq}
\end{equation}
where the sum runs over all admissible families of standard wall in $\bar{V}$.
In this expression the interface has been rewritten in terms of a gas of walls
and thus can be viewed as a model over a two--dimensional lattice. The second
factor in (\ref{qqq}) gives an effective interaction between walls. A theory
of cluster expansions may be developed for this system either directly, as in
Refs. {\cite{BLPO}}, {\cite{BLP2}}, {\cite{BLP3}}, or, equivalently, by
transforming it into a polymer system, as in Ref. {\cite{HKZ}}. This last
method is explained in the two following subsections. 

Again to be able to control this system in terms of convergent cluster
expansion, we need good decaying properties of the activities of walls with
respect to the area $|w| = |w^{\tmop{bk}} | + |w^{W^+} | + |w^{W^-} |$ of the
walls. It follows from easy geometrical observations $( |w| \geqslant 2 ( \pi
( w^{\tmop{bk}} ) + |w^{W^+} | + |w^{W^-} | )$ that:
\begin{equation}
  \rho ( w ) \leqslant e^{- \beta ( J - |K| ) |w|} \label{bmur}
\end{equation}

For $\tilde{Q} ( \bar{V}', \beta )$, we define the walls of the interfaces
$\tilde{I}$ in the same way, getting:
\begin{equation}
  \tilde{Q} ( \bar{V}', \beta ) = e^{- 4 \beta JLN} \sum_{\{ w_1, \ldots, w_n
  \}_{\tmop{adm}} \subset \bar{V}'} \prod_{i = 1}^n \tilde{\rho} ( w_i ) \exp
  \{ - \sum_{X \nsim \tilde{I}, X \subset \bar{V}'} \Phi ( X ) \} \label{qqqq}
\end{equation}
where the activities of walls are defined by:
\begin{equation}
  \tilde{\rho} ( w ) \text{$= \exp \{ - 2 \beta J ( |w| - | \pi ( w ) | ) \}$}
  \label{rrrr}
\end{equation}

\subsection{Decorated interfaces\label{de}}

We are going to rewrite the sum of the R.H.S. of (\ref{qqq}) and (\ref{qqqq})
as a sum of certain elements, which we call {\tmem{decorated interfaces}}. For
$Q ( \bar{V}, \beta )$ the decorated interfaces are defined as the quadruplets
$I_{\tmop{de}} = ( I,\mathcal{D},\mathcal{D}^+,\mathcal{D}^- )$, where $I$ is
an interface and $\mathcal{D},\mathcal{D}^+,\mathcal{D}^-$ are finite set of
clusters imcompatible with the interface.

Given an interface $I$, or what is the same, an admissible family of standard
walls $\{ w_1, \ldots, w_n \}$ such that $I = I ( w_1, \ldots, w_n )$, we
consider the term $B ( I ( w_1, \ldots, w_n ), \bar{V} )$ and define the
weight factors:
\begin{eqnarray*}
  \hat{\psi} ( X ) & = & e^{- \Phi ( X )} - 1\\
  \hat{\psi}_K^{\pm} ( X ) & = & e^{- \Phi_K^{\pm} ( X )} - 1
\end{eqnarray*}
We next define {\tmem{decorations}} $D$ as connected sets of clusters in $V$.
A decoration $D$ is obviously a cluster and there are three types of
decorations for which either all the clusters of the decoration are compatible
with the substrate or all these clusters are incompatible with $W^+$, or they
are incompatible with $W^-$. We define their weights by: 
\begin{eqnarray}
  \psi ( D ) & = & \prod_{X \in D} \hat{\psi} ( X ) \\
  \psi^{\pm}_K ( D ) & = & \prod_{X \in D} \hat{\psi}^{\pm}_K ( X ) 
\end{eqnarray}
and let $\tmop{supp} D = \cup_{X \in D} X$ denotes the support of the family
$D$, and let $|D| = \sum_{X \in D} |X|$ denotes its area. Then,
\begin{eqnarray*}
  e^{B_K ( I ( w_1, \ldots, w_n ), \bar{V} )} & = &
  \prod_{\tmscript{\begin{array}{l}
    X \nsim I\\
    X \sim W
  \end{array}}} [ 1 + \hat{\psi} ( X ) ] \prod_{\tmscript{\begin{array}{l}
    X \nsim I\\
    X \nsim W^+
  \end{array}}} [ 1 + \hat{\psi}_K^+ ( X ) ] \prod_{\tmscript{\begin{array}{l}
    X \nsim I\\
    X \nsim W^-
  \end{array}}} [ 1 + \hat{\psi}_K^- ( X ) ]\\
  & = & \sum_{\tmscript{\begin{array}{l}
    \{ D_1, \ldots, D_n \}_{\tmop{comp}}\\
    D_i \nsim I, D_i \sim W
  \end{array}}} \prod_{i = 1}^n \psi^{} ( D_i )
  \sum_{\tmscript{\begin{array}{l}
    \{ D_1, \ldots, D_m \}_{\tmop{comp}}\\
    D_i \nsim I, D_i \sim W^+
  \end{array}}} \prod_{i = 1}^m \psi^+_K ( D_i )\\
  &  & \times \sum_{\tmscript{\begin{array}{l}
    \{ D_1, \ldots, D_k \}_{\tmop{comp}}\\
    D_i \nsim I, D_i \sim W^-
  \end{array}}} \prod_{i = 1}^k \psi^-_K ( D_i )
\end{eqnarray*}

Starting from (\ref{qqq}), these definitions lead to the expression of
$\text{$\text{$\text{$Q ( \bar{V}, \beta )$}$}$}$ as a sum over the
above--mentioned quadruplets, namely:
\begin{eqnarray}
  \text{$\text{$Q ( \bar{V}, \beta )$}$} & = & e^{\ell ( \bar{V} ) - 2 \beta
  JL_2 L_3} \sum_{\{ w_1, \ldots, w_n \}_{\tmop{comp}} \subset \bar{V}}
  \prod_{i = 1}^n \rho ( w_i ) \sum_{\tmscript{\begin{array}{l}
    \mathcal{D}= \{ D_1, \ldots, D_n \}_{\tmop{comp}} \subset \bar{V}\\
    D_i \nsim I, D_i \sim W
  \end{array}}} \prod_{i = 1}^n \psi ( D_i ) \nonumber\\
  &  & \times \sum_{\tmscript{\begin{array}{l}
    \mathcal{D}= \{ D_1, \ldots, D_m \}_{\tmop{comp}} \subset \bar{V}\\
    D_i \nsim I, D_i \nsim W^+
  \end{array}}} \prod_{i = 1}^m \psi^+_K ( D_i )
  \sum_{\tmscript{\begin{array}{l}
    \mathcal{D}= \{ D_1, \ldots, D_k \}_{\tmop{comp}} \subset \bar{V}\\
    D_i \nsim I, D_i \nsim W^-
  \end{array}}} \prod_{i = 1}^k \psi^-_K ( D_i ) 
\end{eqnarray}
Note that under the hypothesis of Lemma \ref{l:cluster}, the weights of
decorations may be bounded as follows:
\begin{eqnarray}
  | \psi ( D ) | & \leqslant & e - 1 )^{\| D \|} \prod_{X \in D} | \Phi ( X )
  | \leqslant ( e - 1 )^{\| D \|} e^{- [ 2 \beta J - a_1 ] |D|} 
  \label{bdec1}\\
  | \psi^{\pm} ( D ) | & \leqslant & e - 1 )^{\| D \|} \prod_{X \in D} |
  \Phi^{\pm} ( X ) \leqslant ( e - 1 )^{\| D \|} e^{- [ \beta ( J \pm K ) -
  a_1 ] |D|}  \label{bdec2}
\end{eqnarray}
where $\| D \|$ is the number of clusters of the decoration $D$. This follows
from the fact that $|e^x - 1| \leqslant ( e - 1 ) |x|$ for $|x| \leqslant 1$.

For $\tilde{Q} ( \bar{V}', \beta )$, the decorated interfaces are couples,
$\text{$\tilde{I}_{\tmop{de}} = ( \tilde{I},\mathcal{D})$}$, where $\tilde{I}$
is an interface and $\mathcal{D}$ is a finite set of clusters in $\bar{V}'$
incompatible with the interface $\tilde{I}$. The same analysis leads to
\begin{equation}
  \tilde{Q} ( \bar{V}', \beta ) = e^{- 4 \beta JL_2 L_3} \sum_{\{ w_1, \ldots,
  w_n \}_{\tmop{comp}} \subset \bar{V}'} \prod_{i = 1}^n \tilde{\rho} ( w_i )
  \sum_{\tmscript{\begin{array}{l}
    \mathcal{D}= \{ D_1, \ldots, D_n \}_{\tmop{comp}} \subset \bar{V}'\\
    D_i \nsim \tilde{I}
  \end{array}}} \prod_{i = 1}^n \psi ( D_i )
\end{equation}

\subsection{Aggregates\label{agg}}

Let $I_{\tmop{de}} = ( I,\mathcal{D},\mathcal{D}^+,\mathcal{D}^- )$ be a
decorated interface. A quadruplet $\alpha =
(\mathfrak{W},\mathfrak{D},\mathfrak{D}^+,\mathfrak{D}^- )$ where
$\mathfrak{W}$ is a subset of the set of walls $\mathcal{W}( I )$ of the
interface, $\mathfrak{D}$ is a subset of $\mathcal{D}$ and
$\mathfrak{D}^{\pm}$ are subsets of $\mathcal{D}^{\pm}$, is called an
{\tmem{aggregate}}, if its projection $\pi ( a ) \assign \pi (\mathfrak{W})
\cup \pi (\mathfrak{D}) \cup \pi (\mathfrak{D}^+ ) \cup \pi (\mathfrak{D}^- )$
on the plane $P$ is a connected set (in $\mathbb{R}^2$). If there exists a
decorated interface $I_{\tmop{de}}$ such that $\alpha$ is the unique aggregate
of $I_{\tmop{de}}$, $\alpha$ is called a {\tmem{standard aggregate}} (or
{\tmem{aggregate in standard position}}). We observe that the following
geometrical property holds: for any aggregate $\alpha$, there is a standard
aggregate which is just a translate of $\alpha$ (with respect to the
$i_1$--axis). A set of standard aggregates with paiwise disjoint projections
is called an {\tmem{admissible}} family.

Given a decorated interface $I_{\tmop{de}} = (
I,\mathcal{D},\mathcal{D}^+,\mathcal{D}^- )$, one says that $\alpha$ is an
aggreagate of $I_{\tmop{de}}$ if the projection $\pi ( \alpha )$ is a
connected component of $\pi (\mathcal{W}( I ) ) \cup \pi (\mathcal{D}) \cup
\pi ( D^+ ) \cup \pi (\mathcal{D}^- )$. The mapping that associates with a
decorated interface its aggregates in standard position is a bijection onto
the admissible family of standard aggregates.

We define the activity of an aggregate $\text{$\alpha =
(\mathfrak{W},\mathfrak{D},\mathfrak{D}^+,\mathfrak{D}^- )$}$ by:
\begin{equation}
  \omega ( \alpha ) = \prod_{w \in \mathfrak{W}} \rho_{} ( w ) \prod_{D \in
  \mathfrak{D}} \psi ( D ) \prod_{D^+ \in \mathfrak{D}^+} \psi^+ ( D^+ )
  \prod_{D^+ \in \mathfrak{D}^-} \psi^- ( D^- ) \label{www}
\end{equation}
We can then express $\text{$Q ( \bar{V}, \beta$)}$ as the following sum (up to
a prefactor) over all admissible families of standard aggregates in $\bar{V}$:
\begin{equation}
  \text{$Q ( \bar{V}, \beta$)=} e^{\ell ( \bar{V} ) - 2 \beta JLN} \sum_{\{
  \alpha_1, \ldots, \alpha_n \}_{\tmop{adm}} \subset \bar{V}} \prod_{i = 1}^n
  \omega ( \alpha_i ) \label{aggregats}
\end{equation}

We say that an aggregate $\alpha =
(\mathfrak{W},\mathfrak{D},\mathfrak{D}^+,\mathfrak{D}^- )$ do not touch the
substrate if there are no wall of $\mathfrak{W}$ touching the substrate and
$\mathfrak{D}^+ =\mathfrak{D}^- = \emptyset$. It is clear from the definitions
(\ref{rrr}) and (\ref{www}) that the activity of aggregates not touching the
substrate do not depend on $K$. We use $\tmop{supp} \alpha = \cup_{w \in
\mathfrak{W}} w \cup_{D \in \mathfrak{D} \cup \mathfrak{D}^+ \cup
\mathfrak{D}^-} D$ to denote the support of the aggregate $\alpha =
(\mathfrak{W},\mathfrak{D},\mathfrak{D}^+,\mathfrak{D}^- )$ and use $| \alpha
| = \sum_{w \in \mathfrak{W}} |w| + \sum_{D \in \mathfrak{D} \cup
\mathfrak{D}^+ \cup \mathfrak{D}^-} |D|$ to denote its area.

Let $\mathcal{L}$ be the line with endpoints $( 1 / 2, 1 / 2, 1 / 2 )$ and $(
1 / 2, L + 1 / 2, 1 / 2 )$, (it lies on the plane between the substrate and
the first layer and on the plane between $W^+$ and $W^-$), and let $\tmop{pr}
( \cdot )$ be the orthogonal projection on the plane $i_3 = 1 / 2$. We notice
that if an aggregate in standard position $\alpha$ touches the substrate, then
necessarily the projection of its support $\tmop{pr} ( \tmop{supp} \alpha )$
contains at least of bond (unit segment) of the line $\mathcal{L}$; we will
write pr(supp $\alpha ) \nsim \mathcal{L}$. Let us remark that the support of
an aggregate in standard position touching the substrate and with no
decoration touching the substrate, contains necessarily a bond of the line
$\mathcal{L}$.

We next introduce the notion of {\tmem{elementary aggregates}} or
{\tmem{elementary walls}}. An aggregate $\alpha =
(\mathfrak{W},\mathfrak{D},\mathfrak{D}^+,\mathfrak{D}^- )$ is called
elementary if it contains only one wall $w$ and no decorations,
$\mathfrak{D}=\mathfrak{D}^+ =\mathfrak{D}^- = \emptyset$, and if the wall $w$
contains $4$ plaquettes, one of them separating a pair between the wall and
the bulk. Obviously, there are two kinds of elementary aggregates, depending
whether the wall $w$ separates a pair between the bulk and $W^+$ or a pair
between the bulk and $W^-$. In the first case the activity of an elementary
aggregate, say $\alpha_{\tmop{el}}^+$, is given by:
\begin{equation}
  \omega ( \alpha^+_{\tmop{el}} ) = e^{- 6 \beta J - 2 \beta K} \label{aggel+}
\end{equation}
while in the second case, for an elementary aggregate separating the bulk and
$W^-$, say $a_{\tmop{el}}^-$, one has:
\begin{equation}
  \omega ( \alpha^-_{\tmop{el}} ) = e^{- 6 \beta J + 2 \beta K} \label{aggel-}
\end{equation}

We observe also that the activity of any aggregate $\alpha =
(\mathfrak{W},\mathfrak{D},\mathfrak{D}^+,\mathfrak{D}^- )$ satisfies the
bound (see (\ref{rrr}), (\ref{bmur}), (\ref{bdec1}), (\ref{bdec2})):
\begin{eqnarray}
  | \omega ( \alpha ) | & \leqslant & \prod_{w \in \mathfrak{W}} e^{- \beta (
  J - |K| ) |w|} \prod_{D \in \mathfrak{D}} | \psi ( D ) | \prod_{D \in
  \mathfrak{D}^+} | \psi^+ ( D ) | \prod_{D \in \mathfrak{D}^-} | \psi ( D^- )
  | \nonumber\\
  & \leqslant & \prod_{w \in \mathfrak{W}} e^{- \beta ( J - |K| ) |w|}
  \prod_{D \in \mathfrak{D}} ( e - 1 )^{\| D \|} e^{- ( 2 \beta J - a_1 ) |D|}
  \nonumber\\
  &  & \times \prod_{D \in \mathfrak{D}^+} ( e - 1 )^{\| D \|} e^{- [ \beta (
  J - K ) - a_1 ] |D|} \prod_{D \in \mathfrak{D}^-} ( e - 1 )^{\| D \|} e^{- [
  \beta ( J + K ) - a_1 ] |D|} \nonumber\\
  & \leqslant & e^{- [ \beta ( J - |K| ) + a_2 ] | \alpha |}  \label{bagg}
\end{eqnarray}
where $a_2 = a_1 + \log ( e - 1 ) = \kappa_{\tmop{cl}} + \log \nu + \log ( e -
1 )$.

This allows to exponentiate the sum in the R.H.S. of (\ref{aggregats}) as
developed in the next subsection.

For $\tilde{Q} ( \bar{V}', \beta )$, and the decorated interface
$\tilde{I}_{\tmop{de}} = ( \tilde{I},\mathcal{D})$, we define the aggregates
as couples $\tilde{\alpha} = (\mathfrak{W},\mathfrak{D})$, where
$\mathfrak{W}$ is a subset of the set of walls $\mathcal{W}( \tilde{I} )$ of
the interface and  $\mathfrak{D}$ is a subset of $\mathcal{D}$. Defining the
avtivities of aggregates by
\begin{equation}
  \tilde{\omega} ( \tilde{\alpha} ) = \prod_{w \in \mathfrak{W}} 
  \tilde{\rho}_{} ( \tilde{\omega} ) \prod_{D \in \mathfrak{D}} \psi ( D )
  \label{wwww}
\end{equation}
one has
\begin{equation}
  \tilde{Q} ( \bar{V}', \beta ) = e^{- 4 \beta JL_2 L_3} \sum_{\{
  \tilde{\alpha}_1, \ldots, \tilde{\alpha}_n \}_{\tmop{adm}} \subset \bar{V}'}
  \prod_{i = 1}^n \tilde{\omega} ( \tilde{\alpha}_i )
\end{equation}

\subsection{Multi-indexes of aggregates}

To exponentiate the sum in the R.H.S. of (\ref{aggregats}), we define, as it
was done for contours, multi-indexes of aggregates. A multi-index (of
aggregate) $Y$ is a function from the set of aggregates into the set of
nonnegative integers. We let $\op{\tmop{supp}} Y = \cup_{\alpha : Y ( \alpha )
\geq 1} \tmop{supp} \alpha$ denotes the support of the multi-index Y. We
define the truncated functional associated with the activity (\ref{www}) of
aggregates by:
\begin{equation}
  \Psi ( Y ) = \frac{a ( Y )}{\prod_{\alpha} Y ( \alpha ) !} \prod_a \omega_{}
  ( \alpha )^{Y ( \alpha )}
\end{equation}
where the combinatorial factor $a ( Y )$ is defined as in (\ref{multi}). 

Again, $a ( Y ) = 0$ and hence $\Psi ( Y ) = 0$ unless $\tmop{supp} Y$ is a
connected set, and whenever $Y$ contains only one aggregate $\alpha$, then
$\Psi_{} ( Y ) = \omega_{} ( \alpha )$.

We say that a multi-index $Y$ do not touch the substrate,  if it is  supported
by aggregates not touching the substrate.

Note that the projection of support of multi-indexes $Y$ touching the
substrate, $\tmop{pr} ( \tmop{supp} Y )$, contains (as the projection of  the
support of standard aggregates touching the substrate) at least a bond of the
line $\mathcal{L}$. We write for such multi-indexes $\tmop{pr} ( \tmop{supp} Y
) \nsim \mathcal{L}$.

A consequence of the previous definitions is that the sum in the R.H.S. of
(\ref{aggregats}) can be exponentiated as a sum over multi--indexes of
aggregates and we get:
\begin{equation}
  \ln \text{$Q ( \bar{V}, \beta$)=} \ell ( \bar{V} ) - 2 \beta JL_2 L_3 +
  \sum_{Y \subset \bar{V}} \Psi ( Y ) \label{logq}
\end{equation}
where the sum runs over (non-empty) multi-indexes of aggregates $Y$ with
support in $\bar{V}$.

\begin{lemma}
  \label{l:agg}Assume $\beta ( J - |K| ) \geqslant 2 \kappa_{\tmop{cl}} + \log
  ( 5 \nu ) + \log ( e - 1 )$, and let $\alpha_0$ be a given wall or a given
  contour, then
  \begin{equation}
    \sum_{Y : Y ( \alpha_0 ) \geqslant 1} | \Psi ( Y ) | \leqslant e^{- [
    \beta ( J - |K| ) - a_2 - a_0 ] | \alpha_0 |} = e^{- [ \beta ( J - |K| ) -
    k_{\tmop{cl}} - \log \nu - \log ( e - 1 ) - a_0 ] | \alpha_0 |}
  \end{equation}
  and the series $\sum_{Y : \tmop{supp} Y \ni i} | \Psi ( Y ) |$ is absolutely
  convergent.
\end{lemma}

\begin{proof}
  We first notice that the number of aggregates $\alpha$ of area $| \alpha | =
  n$ is less than $( 4 \nu )^n$. Indeed, an aggregate of area $n$ containing
  at most $2 n$ vertices, the factor $2^{2 n} \geqslant \binom{2 n}{k}$ bounds
  the number of choice of vertices connecting the contours and walls, and the
  factor $\nu^n$ bounds the entropies of contours and walls.
  
  Then, as in Lemma \ref{l:cluster}, we know from Ref. {\cite{M}}, that under
  the convergence condition
  \begin{equation}
    \omega ( \alpha ) \leqslant ( e^{\mu ( \alpha )} - 1 ) \exp [ -
    \sum_{\alpha \nsim \alpha_0} \mu ( \alpha ) ]
  \end{equation}
  (where $\mu$ is a positive function), then
  \begin{equation}
    \sum_{Y : Y ( \alpha_0 ) \geqslant 1} | \Psi ( Y ) | \leqslant \mu (
    \alpha_0 )
  \end{equation}

  We choose $\mu ( \alpha ) = e^{- [ \beta ( J - |K| ) - a_2 - a ] | \alpha
  |}$ to get, by taking into account the above remark on the entropy of
  aggregates and that the minimal area of walls (and thus of aggregates) is
  $4$, that
  \begin{equation}
    \sum_{\alpha \nsim \alpha_0} \mu ( \alpha_0 ) \leqslant 2| \alpha |
    \sum_{n \equallim 4}^{\infty} ( 4 \nu )^n e^{- n [ \beta ( J - |K| ) - a_2
    - a ]} \leqslant \frac{1}{e^{\beta ( J - |K| ) - a_2 - a - \log ( 4 \nu )}
    - 1} | \gamma |
  \end{equation}
  provided $\text{$2 ( 4 \nu )^3 e^{- 3 [ \beta ( J - |K| ) - a_2 - a ]}
  \leqslant 1$}$. The factor $2$ stems from the fact that an aggregate of area
  $| \alpha |$ contains at most $2| \alpha |$ vertices. Taking into account
  the bound (\ref{bagg}) on the activities of aggregates and using $\mu (
  \gamma ) \leqslant e^{\mu ( \gamma )} - 1$, one sees that the convergence
  condition is satisfied whenever
  \begin{equation}
    \beta ( J - |K| ) \geqslant \log ( 4 \nu ) + a_2 + a + \log \frac{a +
    1}{a}
  \end{equation}
  Again we take the value $a = a_0 = \frac{\sqrt{5} - 1}{2}$ that minimizes
  the function $a + \log \frac{a + 1}{a}$: its provides the condition given in
  the lemma.
\end{proof}

We also consider the multi--indexes of the aggregates $\tilde{\alpha}$ and
define the truncated functional associated with the activity (\ref{wwww}) of
these aggregates by:
\begin{equation}
  \tilde{\Psi} ( Y ) = \frac{a ( Y )}{\prod_{\tilde{\alpha}} Y (
  \tilde{\alpha} ) !} \prod_{\tilde{\alpha}}  \tilde{\omega}_{} (
  \tilde{\alpha} )^{Y ( \tilde{\alpha} )}
\end{equation}
The quantity $\ln \tilde{Q} ( \bar{V}', \beta )$ can be written as the
following sum:
\begin{equation}
  \ln \tilde{Q} ( \bar{V}', \beta ) = - 4 \beta JL_2 L_3 + \sum_{Y \subset
  \bar{V}'}  \tilde{\Psi} ( Y ) \label{logqq}
\end{equation}
Note that for multi--indexes entering in the expression (\ref{logq}) and not
touching the substrate, one has $\Psi ( Y ) = \tilde{\Psi} ( Y )$.

We let $\bar{V}'' = \bar{V}' \setminus \bar{V}$ and $P'$ be the plane $i_3 =
1 / 2$. Notice that the box $\bar{V}''$ is the image of $\bar{V}$ by the
reflection with respect to this plane. We decompose the sum in the R.H.S. of
(\ref{logqq}) as follows:
\begin{equation}
  \sum_{Y \subset \bar{\Lambda}}  \tilde{\Psi} ( Y ) =
  \sum_{\tmscript{\begin{array}{c}
    Y \subset \bar{V}\\
    \tmop{supp} Y \cap P' = \emptyset
  \end{array}}}  \tilde{\Psi} ( Y ) + \sum_{\tmscript{\begin{array}{c}
    Y \subset \bar{V}''\\
    \tmop{supp} Y \cap P' = \emptyset
  \end{array}}}  \tilde{\Psi} ( Y ) + \sum_{\tmscript{\begin{array}{c}
    Y \subset \bar{V}'\\
    \tmop{supp} Y \cap P' \neq \emptyset
  \end{array}}}  \tilde{\Psi} ( Y )
\end{equation}
Obviously the first two sums are equal and the first one can be written as a
sum over multi--indexes in $\bar{V}$ not touching the substrate, we write $Y
\sim W$, giving
\begin{equation}
  \ln \tilde{Q} ( \bar{V}', \beta ) = - 4 \beta JL_2 L_3 + 2
  \sum_{\tmscript{\begin{array}{c}
    Y \subset \bar{V}\\
    Y \sim W
  \end{array}}}  \tilde{\Psi} ( Y ) + \sum_{\tmscript{\begin{array}{c}
    Y \subset \bar{V}'\\
    \tmop{supp} Y \cap P' \neq \emptyset
  \end{array}}} \tilde{\Psi} ( Y ) \label{logqqq}
\end{equation}

\subsection{Existence of the line tension and proof of Theorem \ref{T1}}

We start from equations (\ref{logq}) and (\ref{logqqq}) and take into account
that for multi-indexes $Y$ not touching the substrate, $\Psi ( Y ) =
\tilde{\Psi} ( Y )$, to get
\begin{eqnarray}
  \ln \frac{\text{$Q ( \bar{V}, \beta$)}}{( \tilde{Q} ( \bar{V}', \beta ) )^{1
  / 2}} & = & \ell ( \bar{V} ) + \sum_{\tmscript{\begin{array}{c}
    Y \subset \bar{V}\\
    Y \nsim W
  \end{array}}} \Psi ( Y ) - \frac{1}{2} \sum_{\tmscript{\begin{array}{c}
    Y \subset \bar{V}'\\
    \tmop{supp} Y \cap P' \neq \emptyset
  \end{array}}}  \tilde{\Psi} ( Y ) \nonumber\\
  & = & \ell ( \bar{V} ) + \sum_{\tmscript{\begin{array}{c}
    Y \subset \bar{V}\\
    \tmop{pr} ( \tmop{supp} Y ) \nsim \mathcal{L}
  \end{array}}} \Psi ( Y ) - \frac{1}{2} \sum_{\tmscript{\begin{array}{c}
    Y \subset \bar{V}'\\
    \tmop{supp} Y \cap P' \neq \emptyset
  \end{array}}}  \tilde{\Psi} ( Y ) 
\end{eqnarray}
where the first sum in the R.H.S. of the above equation is over multi-indexes
of aggregates with support in $\bar{V}$  touching the substrate: recall that
the projection of the support of multi--indexes of aggregates touching the
substrate contains at least a bond of the line $\mathcal{L}$. Notice that the
(orthogonal ) projection $\tmop{pr} ( \tmop{supp} Y )$ of support of
muti--indexes $Y$ whose support intersects the plane $P'$ contains necessarily
a bond of the Line $\mathcal{L}$. We write $\tmop{pr} ( \tmop{supp} Y ) \nsim
\mathcal{L}$ for such clusters, getting:
\[ \ln \frac{\text{$Q ( \bar{V}, \beta$)}}{( \tilde{Q} ( \bar{V}', \beta )
   )^{1 / 2}} = \begin{array}{ll}
     = & \ell ( \bar{V} ) + \sum_{\tmscript{\begin{array}{c}
       Y \subset \bar{V}\\
       \tmop{pr} ( \tmop{supp} Y ) \nsim \mathcal{L}
     \end{array}}} \Psi ( Y ) - \frac{1}{2} \sum_{\tmscript{\begin{array}{c}
       Y \subset \bar{V}'\\
       \tmop{pr} ( \tmop{supp} Y ) \nsim \mathcal{L}
     \end{array}}}  \tilde{\Psi} ( Y ) \eqnumber
   \end{array} \]
Let us introduce the infinite boxes
\[ \hat{V} = \lim_{L_3 \rightarrow \infty}  \bar{V} = \{ i = ( i_1, i_2, i_3
   ) \in \mathbb{L} : i_1 \in \mathbb{Z}, 0 \leq i_2 \leq L, 1 \leq i_3 \leq
   \infty \} \]
and $\hat{V}' = \lim_{L_3 \rightarrow \infty}  \bar{V}'$. The absolute
convergence of the series of truncated functions (of clusters and
multi-indexes of aggregates) implies the existence of the limit, $\lim_{L_3
\rightarrow \infty} \log [ \text{$Q ( \bar{V}, \beta$)} / ( \tilde{Q} (
\bar{V}', \beta ) )^{1 / 2} ]$. We denote this limit by $F_{} ( \hat{V} )$.
One has
\begin{equation}
  F ( \hat{V} ) = \ell ( \hat{V} ) + \sum_{\tmscript{\begin{array}{c}
    Y \subset \hat{V}\\
    \tmop{pr} ( \tmop{supp} Y ) \nsim \mathcal{L}
  \end{array}}} \Psi ( Y ) - \frac{1}{2} \sum_{\tmscript{\begin{array}{c}
    Y \subset \hat{V}'\\
    \tmop{pr} ( \tmop{supp} Y ) \nsim \mathcal{L}
  \end{array}}}  \tilde{\Psi} ( Y )
\end{equation}
Here $\ell ( \hat{V} )$ is defined by (\ref{aaa}) with $V$ replaced by
$\hat{V}$, and the first sum in the R.H.S. is over multi-indexes of aggregates
with support in $\hat{V}$ touching the substrate.

We denote $\tmmathbf{L}$ the infinte line $( i_1 = 1 / 2, i_3 = 1 / 2 )$.
Then, again the absolute convergence of the series of truncated functions
implies the existence of the limit $\lim_{L_2 \rightarrow \infty} F ( \hat{V}
) / L_2$. As a result, we get:
\begin{equation}
  \label{lt} - \beta \lambda ( \beta ) = s_{\tmop{cl}} + s_{\tmop{agg}} +
  s_{\tmop{aggr}}
\end{equation}
where the convergent series $s_{\tmop{cl}}$, $s_{\tmop{agg}}$, and
$s_{\tmop{aggr}}$ are given by:

\begin{eqnarray}
  s_{\tmop{cl}} & = & \frac{1}{2} \sum_{\tmscript{\begin{array}{c}
    X : X \subset \mathbb{L}\\
    \tmop{pr} ( \tmop{supp} X ) \ni b
  \end{array}}} \frac{\Phi^+ ( X ) + \Phi^- ( X )}{| \tmop{pr} ( \tmop{supp} X
  ) \cap \tmmathbf{L}|}  \label{scl}\\
  s_{\tmop{agg}} & = & \sum_{\tmscript{\begin{array}{c}
    Y : \tmop{supp} Y \subset \mathbb{L}\\
    \tmop{pr} ( \tmop{supp} Y ) \ni b
  \end{array}}} \frac{\Psi_{} ( Y )}{| \tmop{pr} ( \tmop{supp} Y ) \cap
  \tmmathbf{L}|}  \label{sagg}\\
  s_{\tmop{aggr}} & = & - \frac{1}{2} \sum_{\tmscript{\begin{array}{c}
    Y : \tmop{supp} Y \subset \mathbb{Z}^3\\
    \tmop{pr} ( \tmop{supp} Y ) \ni b
  \end{array}}}  \frac{\tilde{\Psi}_{} ( Y )}{| \tmop{pr} ( \tmop{supp} Y )
  \cap \tmmathbf{L}|}  \label{saggr}
\end{eqnarray}
The sum $s_{\tmop{cl}}$ is over clusters inside the semi-infinite lattice
$\mathbb{L}$ whose projection on the plane $i_3 = 1 / 2$ contains a bond $b$
of the line $\tmmathbf{L}$, and the sums $s_{\tmop{agg}}$ and $s_{\tmop{aggr}}
$are over multi-indexes of aggregates with support inside $\mathbb{L}$ and
$\mathbb{Z}^3$ respectively, whose projection contains a bond $b$ of the line
$\tmmathbf{L}$, $| \tmop{pr} ( \tmop{supp} X ) \cap \tmmathbf{L}|$ and $|
\tmop{pr} ( \tmop{supp} Y ) \cap \tmmathbf{L}|$ are the respective lengths of
$\text{$\tmop{pr} ( \tmop{supp} X ) \cap \tmmathbf{L}$}$ and $\tmop{pr} (
\tmop{supp} Y ) \cap \tmmathbf{L}$.

We have used the (standard) decompositions:
\begin{eqnarray*}
  \frac{1}{L_2} \sum_{\tmscript{\begin{array}{l}
    X \subset \hat{V}\\
    X \nsim W^+\\
    X \nsim W^-
  \end{array}}} \Phi^{\pm} ( X ) & = & \frac{1}{L_2} 
  \sum_{\tmscript{\begin{array}{l}
    X \cap \hat{V} \neq \emptyset\\
    X \nsim W^+\\
    X \nsim W^-
  \end{array}}} \Phi^{\pm} ( X ) - \frac{1}{L_2} 
  \sum_{\tmscript{\begin{array}{c}
    X \cap \hat{V} \neq \emptyset, X \cap (\mathbb{L} \setminus \hat{V} )
    \neq \emptyset\\
    X \nsim W^+\\
    X \nsim W^-
  \end{array}}} \Phi^{\pm} ( X )\\
  \frac{1}{L_2} \sum_{\tmscript{\begin{array}{c}
    Y : \tmop{supp} Y \subset \hat{V}\\
    \tmop{pr} ( \tmop{supp} Y ) \nsim \mathcal{L}
  \end{array}}} \Psi ( Y ) & = & \frac{1}{L_2}
  \sum_{\tmscript{\begin{array}{c}
    Y : \tmop{supp} Y \cup \hat{V} \neq \emptyset\\
    \tmop{pr} ( \tmop{supp} Y ) \nsim \mathcal{L}
  \end{array}}} \Psi ( Y ) - \frac{1}{L_2} \sum_{\tmscript{\begin{array}{c}
    Y : \tmop{supp} Y \cup \hat{V} \neq \emptyset\\
    \tmop{supp} Y \cup (\mathbb{L} \setminus \hat{V} ) \neq \emptyset\\
    \tmop{pr} ( \tmop{supp} Y ) \nsim \mathcal{L}
  \end{array}}} \Psi ( Y )\\
  \frac{1}{L_2} \sum_{\tmscript{\begin{array}{c}
    Y \subset \hat{\Omega}\\
    \tmop{pr} ( \tmop{supp} Y ) \nsim \mathcal{L}
  \end{array}}}  \tilde{\Psi} ( Y ) & = & \frac{1}{L_2}
  \sum_{\tmscript{\begin{array}{c}
    Y : \tmop{supp} Y \cap \hat{V}' \neq \emptyset\\
    \tmop{pr} ( \tmop{supp} Y ) \nsim \mathcal{L}
  \end{array}}} \tilde{\Psi} ( Y ) - \frac{1}{L_2}
  \sum_{\tmscript{\begin{array}{c}
    Y : \tmop{supp} Y \cup \hat{V}' \neq \emptyset\\
    \tmop{supp} Y \cup (\mathbb{Z}^3 \setminus \hat{V}' ) \neq \emptyset\\
    \tmop{pr} ( \tmop{supp} Y ) \nsim \mathcal{L}
  \end{array}}} \tilde{\Psi} ( Y )
\end{eqnarray*}
Both the second sums in the R.H.S. of the above equations tends to $0$ in the
limit $L_2 \rightarrow \infty$, while the first sums (in these R.H.S.) can be
rewritten, by taking into account the translation invariance of $\Phi^{\pm} (
X )$, $\Psi ( Y )$, and $\tilde{\Psi} ( Y )$ with respect to the
$i_2$--direction, as:
\begin{eqnarray*}
  \sum_{\tmscript{\begin{array}{l}
    X \cap \hat{V} \neq \emptyset\\
    X \nsim W^+\\
    X \nsim W^-
  \end{array}}} \Phi^{\pm} ( X ) & = & \sum_{b \in \mathcal{L}}  \sum_{X :
  \tmop{pr} ( \tmop{supp} X ) \ni b}  \frac{\Phi^{\pm} ( X )}{| \tmop{pr} (
  \tmop{supp} X ) \cap \mathcal{L}|}\\
  \sum_{\tmscript{\begin{array}{c}
    Y : \tmop{supp} Y \cup \hat{V} \neq \emptyset\\
    \tmop{pr} ( \tmop{supp} Y ) \nsim \mathcal{L}
  \end{array}}} \Psi ( Y ) & = & \sum_{b \in \mathcal{L}}  \sum_{Y : \tmop{pr}
  ( \tmop{supp} Y ) \ni b}  \frac{\Psi ( Y )}{| \tmop{pr} ( \tmop{supp} Y )
  \cap \mathcal{L}|}\\
  \sum_{\tmscript{\begin{array}{c}
    Y : \tmop{supp} Y \cap \hat{V}' \neq \emptyset\\
    \tmop{pr} ( \tmop{supp} Y ) \nsim \mathcal{L}
  \end{array}}} \tilde{\Psi} ( Y ) & = & \sum_{b \in \mathcal{L}}  \sum_{Y :
  \tmop{pr} ( \tmop{supp} Y ) \ni b}  \frac{\tilde{\Psi} ( Y )}{| \tmop{pr} (
  \tmop{supp} Y ) \cap \mathcal{L}|}
\end{eqnarray*}
which lead to (\ref{scl}), (\ref{sagg}), and (\ref{saggr}) in the limit $L_3
\rightarrow \infty$.

The leading terms of the three series may be easily found. For the series
$s_{\tmop{cl}}$, we notice that the smallest cluster $X$ entering in the sum
contains $10$ plaquettes among then one plaquette separates the lattice from
$W^+$ and another one separates the lattice from $W^-$. Since, as mentioned
above, the truncated functions of such clusters coincide with the weight of
the corresponding contour $\gamma$ ($| \gamma^{\tmop{bk}} | = 8, | \gamma^W |
= 2$), one has
\begin{equation}
  s_{\tmop{cl}} = \frac{1}{2} ( e^{- 16 \beta J - 4 \beta K} + e^{- 16 \beta J
  + 4 \beta K} ) + \tmop{higher} \tmop{order} \label{scl1}
\end{equation}
For the series $s_{\tmop{agg}}$, the smallest multi--indexes are the ones
corresponding to the elementary aggregates. As noticed above, the truncated
functions of such multi-indexes coincide with the weights of the corresponding
aggregates (see (\ref{aggel+}) and \ref{aggel-})). We have thus
\begin{equation}
  s_{\tmop{agg}} = e^{- 6 \beta J} ( e^{2 \beta K} + e^{- 2 \beta K} ) +
  \tmop{higher} \tmop{order} \label{sagg1}
\end{equation}
Finally, for the series $s_{\tmop{aggr}}$, the smallest cluster entering in
the sum, contains $4$ plaquettes. The truncated function of such multi--index
coincide with the weight (given by (\ref{rrrr}) and (\ref{wwww})) of the
corresponding aggregate, and therefore:
\begin{equation}
  s_{\tmop{aggr}} = e^{- 8 \beta J} + \tmop{higher} \tmop{order}
\end{equation}

{\subsection*{Acknowledgements}}

L.L. thanks the warm hospitality and financial support of the Centre de
Physique Théorique and the Université du Sud Toulon--Var.

\[  \]

\end{document}